\documentclass{pasj00}
\draft

\begin{document}
\SetRunningHead{Murakami et al.}{AKARI Mission}
\Received{2007/05/31}
\Accepted{2007/08/09}

\title{The Infrared Astronomical Mission AKARI
\thanks{AKARI is a JAXA project with the participation of ESA.}}


\author{%
Hiroshi~\textsc{Murakami}\altaffilmark{1},
Hajime~\textsc{Baba}\altaffilmark{1},
Peter~\textsc{Barthel}\altaffilmark{2},
David~L.~\textsc{Clements}\altaffilmark{3},
Martin~\textsc{Cohen}\altaffilmark{4},
Yasuo~\textsc{Doi}\altaffilmark{5},
Keigo~\textsc{Enya}\altaffilmark{1},
Elysandra~\textsc{Figueredo}\altaffilmark{6},
Naofumi~\textsc{Fujishiro}\altaffilmark{1,7}\thanks{Present Address is Cybernet systems Co. Ltd., 
Bunkyo-ku, Tokyo 112-0012, Japan}, 
Hideaki~\textsc{Fujiwara}\altaffilmark{8},
Mikio~\textsc{Fujiwara}\altaffilmark{9},
Pedro~\textsc{Garcia-Lario}\altaffilmark{10},
Tomotsugu~\textsc{Goto}\altaffilmark{1},
Sunao~\textsc{Hasegawa}\altaffilmark{1},
Yasunori~\textsc{Hibi}\altaffilmark{11}\thanks{Present Address is National Astronomical Observatory 
of Japan, National Institutes
of Natural Sciences, Mitaka, Tokyo 181-8588, Japan},
Takanori~\textsc{Hirao}\altaffilmark{11}\thanks{Present Address is Research Institute of Science 
and Technology for Society,  Japan Science and Technology Agency, Kawaguchi, Saitama 332-0012, Japan}, 
Norihisa~\textsc{Hiromoto}\altaffilmark{12},
Seung~Soo~\textsc{Hong}\altaffilmark{13},
Koji~\textsc{Imai}\altaffilmark{1},
Miho~\textsc{Ishigaki}\altaffilmark{1}\thanks{Present Address is Astronomical Institute, 
Tohoku University, Aoba-ku, Sendai 980-77, Japan},
Masateru~\textsc{Ishiguro}\altaffilmark{13},
Daisuke~\textsc{Ishihara}\altaffilmark{8},
Yoshifusa~\textsc{Ita}\altaffilmark{1}\thanks{Present Address is National Astronomical Observatory 
of Japan, National Institutes of Natural Sciences, Mitaka, Tokyo 181-8588, Japan}, 
Woong-Seob~\textsc{Jeong}\altaffilmark{1},
Kyung~Sook~\textsc{Jeong}\altaffilmark{13},
Hidehiro~\textsc{Kaneda}\altaffilmark{1},
Hirokazu~\textsc{Kataza}\altaffilmark{1},
Mitsunobu~\textsc{Kawada}\altaffilmark{11},
Toshihide~\textsc{Kawai}\altaffilmark{14},
Akiko~\textsc{Kawamura}\altaffilmark{11},
Martin~F.~\textsc{Kessler}\altaffilmark{10,15},
Do~\textsc{Kester}\altaffilmark{16},
Tsuneo~\textsc{Kii}\altaffilmark{1},
Dong Chan~\textsc{Kim}\altaffilmark{17},
Woojung~\textsc{Kim}\altaffilmark{1}\thanks{Present Address is Semiconductor Business Group, 
Sony Corporation, 4-14-1 Asahi-cho, Atsugi, Kanagawa 243-0014, Japan},
Hisato~\textsc{Kobayashi}\altaffilmark{1,7},
Bon~Chul~\textsc{Koo}\altaffilmark{13},
Suk~Minn~\textsc{Kwon}\altaffilmark{18},
Hyung~Mok~\textsc{Lee}\altaffilmark{13},
Rosario~\textsc{Lorente}\altaffilmark{10},
Sin'itirou~\textsc{Makiuti}\altaffilmark{1},
Hideo~\textsc{Matsuhara}\altaffilmark{1},
Toshio~\textsc{Matsumoto}\altaffilmark{1},
Hiroshi~\textsc{Matsuo}\altaffilmark{19},
Shuji~\textsc{Matsuura}\altaffilmark{1},
Thomas~G.~\textsc{M\"{u}ller}\altaffilmark{20},
Noriko~\textsc{Murakami}\altaffilmark{11},
Hirohisa~\textsc{Nagata}\altaffilmark{1}, 
Takao~\textsc{Nakagawa}\altaffilmark{1},
Takahiro~\textsc{Naoi}\altaffilmark{1}
\thanks{Present Address is National Astronomical Observatory of Japan, National Institutes of 
Natural Sciences, Mitaka, Tokyo 181-8588, Japan},
Masanao~\textsc{Narita}\altaffilmark{1},
Manabu~\textsc{Noda}\altaffilmark{21},
Sang~Hoon~\textsc{Oh}\altaffilmark{13},
Akira~\textsc{Ohnishi}\altaffilmark{1},
Youichi~\textsc{Ohyama}\altaffilmark{1},
Yoko~\textsc{Okada}\altaffilmark{1},
Haruyuki~\textsc{Okuda}\altaffilmark{1}, 
Sebastian~\textsc{Oliver}\altaffilmark{22}, 
Takashi~\textsc{Onaka}\altaffilmark{8}, 
Takafumi~\textsc{Ootsubo}\altaffilmark{11}, 
Shinki~\textsc{Oyabu}\altaffilmark{1}, 
Soojong~\textsc{Pak}\altaffilmark{23}, 
Yong-Sun~\textsc{Park}\altaffilmark{13}, 
Chris~P.~\textsc{Pearson}\altaffilmark{1,10}, 
Michael~\textsc{Rowan-Robinson}\altaffilmark{3},
Toshinobu~\textsc{Saito}\altaffilmark{1,7},
Itsuki~\textsc{Sakon}\altaffilmark{8}, 
Alberto~\textsc{Salama}\altaffilmark{10}, 
Shinji~\textsc{Sato}\altaffilmark{11},
Richard S.~\textsc{Savage}\altaffilmark{22}
Stephen~\textsc{Serjeant}\altaffilmark{6}, 
Hiroshi~\textsc{Shibai}\altaffilmark{11}, 
Mai~\textsc{Shirahata}\altaffilmark{1}, 
Jungjoo~\textsc{Sohn}\altaffilmark{13}, 
Toyoaki~\textsc{Suzuki}\altaffilmark{1,7}, 
Toshinobu~\textsc{Takagi}\altaffilmark{1}, 
Hidenori~\textsc{Takahashi}\altaffilmark{24}, 
Toshihiko~\textsc{Tanab\'{e}}\altaffilmark{25}, 
Tsutomu~T.~\textsc{Takeuchi}\altaffilmark{26},
Satoshi~\textsc{Takita}\altaffilmark{1,27},
Matthew~\textsc{Thomson}\altaffilmark{22},
Kazunori ~\textsc{Uemizu}\altaffilmark{1}, 
Munetaka~\textsc{Ueno}\altaffilmark{5}, 
Fumihiko~\textsc{Usui}\altaffilmark{1}, 
Eva~\textsc{Verdugo}\altaffilmark{10}, 
Takehiko~\textsc{Wada}\altaffilmark{1}, 
Lingyu~\textsc{Wang}\altaffilmark{3}
Toyoki~\textsc{Watabe}\altaffilmark{14},
Hidenori~\textsc{Watarai}\altaffilmark{1}\thanks{Present Address is 
Office of Space Applications, Japan Aerospace Exploration Agency, Tsukuba, Ibaraki 305-8505, Japan}, 
Glenn~J.~\textsc{White}\altaffilmark{6,28}, 
Issei~\textsc{Yamamura}\altaffilmark{1}, 
Chisato~\textsc{Yamauchi}\altaffilmark{1},
and
Akiko~\textsc{Yasuda}\altaffilmark{1,29}
}
%
\altaffiltext{1}{Institute of Space and Astronautical Science, Japan Aerospace Exploration Agency, 
Sagamihara, Kanagawa 229-8510, Japan}
\email{hmurakam@ir.isas.jaxa.jp}
\altaffiltext{2}{Kapteyn Astronomical Institute, University of Groningen, Groningen, 9700 AV, 
The Netherlands }
\altaffiltext{3}{Blackett Laboratory, Imperial College London, Prince Consort Road, London SW7 2AZ, U.K.}
\altaffiltext{4}{Radio Astronomy Laboratory, 601 Campbell Hall, University of California, Berkeley, 
CA94720, U.S.A.}
\altaffiltext{5}{Department of Earth Science and Astronomy, Graduate School of Arts and Sciences, 
The University of Tokyo, Meguro-ku, Tokyo 153-8902, Japan}
\altaffiltext{6}{Department of Physics and Astronomy, The Open University, Milton Keynes MK7 6AA, U.K.}
\altaffiltext{7}{Department of Physics, Graduate School of Science, The University of Tokyo, 
Bunkyo-ku, Tokyo 113-0033, Japan}
\altaffiltext{8}{Department of Astronomy, Graduate School of Science, The University of Tokyo, 
Bunkyo-ku, Tokyo 113-0033, Japan}
\altaffiltext{9}{Research Department 1, National Institute of Information and Communications 
Technology, Koganei, Tokyo 184-8795, Japan}
\altaffiltext{10}{European Space Agency, ESAC, P.O. Box 78, 28691 Villanueva de la Ca\~nada, 
Madrid, Spain}
\altaffiltext{11}{Division of Particle and Astrophysical Sciences, Graduate School of Science, 
Nagoya University, Furo-cho, Chikusa-ku, Nagoya 464-8602, Japan}
\altaffiltext{12}{Optelectronics and Electromagnetic Wave Engineering, Shizuoka University, 3-1-5 Johoku, Hamamatsu, Japan}
\altaffiltext{13}{Department of Physics and Astronomy, Seoul National University, Shillimdong Kwanak-gu, Seoul 151-747, Korea}
\altaffiltext{14}{Technical Center of Nagoya University, Furo-cho, Chikusa-ku, Nagoya 464-8602, Japan}
\altaffiltext{15}{European Space Agency, ESTEC, Keplerlaan 1, 2200 AG Noordwijk, The Netherlands}
\altaffiltext{16}{Netherlands Institute for Space Research SRON, AV Groningen, Groningen, The Netherlands}
\altaffiltext{17}{Department of Astronomy, University of Maryland, College Park, MD 20742-2421, U.S.A.}
\altaffiltext{18}{Department of Science Education, Kangwon National University, Hyoja-dong, Kangwon-do, Chunchon 200-701, Korea}
\altaffiltext{19}{National Astronomical Observatory of Japan, National Institutes of Natural Sciences, Mitaka, Tokyo 181-8588, Japan}
\altaffiltext{20}{Max-Planck-Institut f\"{u}r extraterrestrische Physik, Giessenbachstra{\ss}e, 85748 Garching, Germany}
\altaffiltext{21}{Nagoya City Science Museum, Sakae, Naka-ku, Nagoya 460-0008, Japan}
\altaffiltext{22}{Astronomy Centre, Department of Physics and Astronomy, University of Sussex, Brighton BN1 9QH, U.K.}
\altaffiltext{23}{Department of Astronomy and Space Science, Kyung Hee University, Seocheon-dong, Giheung-gu, Yongin-si, Gyeonggi-do 446-701, Korea}
\altaffiltext{24}{Gunma Astronomical Observatory, Takayama, Agatsuma, Gunma 377-0702, Japan}
\altaffiltext{25}{Institute of Astronomy, Graduate School of Science, The University of Tokyo, Mitaka, Tokyo 181-0015, Japan}
\altaffiltext{26}{Institute for Advanced Research, Nagoya University, Furo-cho, Chikusa-ku, Nagoya 464-8601, Japan}
\altaffiltext{27}{Department of Earth and Planetary Sciences, Graduate School of Science and Engineering,
Tokyo Institute of Technology, 2-12-1 Ookayama, Meguro-ku, Tokyo, 152-8550, Japan}
\altaffiltext{28}{Space Science \& Technology Department, CCLRC Rutherford Appleton Laboratory, Chilton, Didcot, Oxfordshire OX11 0QX, U.K.}
\altaffiltext{29}{The Graduate University for Advanced Studies, Shonan Village, Hayama, Kanagawa 240-0193, Japan}


\maketitle

\begin{abstract}
AKARI, the first Japanese satellite dedicated to infrared astronomy, was launched on 2006 February 21, 
and started observations in May of the same year. AKARI has a 68.5 cm cooled telescope, 
together with two focal-plane instruments, which survey the sky in six wavelength bands 
from the mid- to far-infrared. 
The instruments also have the capability for imaging and spectroscopy in the wavelength range 
2 -- 180 $\mu$m in 
the pointed observation mode,  occasionally inserted into the continuous survey operation. 
The in-orbit cryogen lifetime is expected to be one and a half years. 
The All-Sky Survey will cover more than 90 percent of the whole sky 
with higher spatial resolution and wider wavelength coverage than that of  the previous
IRAS all-sky survey.
Point source catalogues of the All-Sky Survey will be released to the
astronomical community.
The pointed observations will be used for deep surveys of selected sky areas and 
systematic observations of important astronomical targets. These will become an additional future 
heritage of this mission. 
\end{abstract}

\section{Introduction}

ASTRO-F was planned as a Japanese space mission dedicated to infrared astronomy (\cite{murakami_04}, 
\cite{AKARI_Sb}). 
It was successfully launched on 2006 February 21 (UT) on the M-V-8 rocket,
which was developed by the 
Japan Aerospace Exploration Agency (JAXA), from Uchinoura Space Center (USC). 
It was renamed AKARI after the confirmation of successful insertion of the satellite into the orbit. 

AKARI is the second Japanese space mission to carry out infrared astronomy following 
the Infrared Telescope in Space (IRTS; \cite{murakami_96}) onboard the Space Flyer Unit (SFU).
AKARI is designed as an All-Sky Survey mission in the infrared. 
The primary purpose of the mission is to provide second-generation infrared 
catalogues to better spatial resolution and wider spectral coverage than
the first catalogues by the Infra-Red Astronomy Satellite (IRAS) mission (\cite{IRAS}). AKARI is equipped 
with a cryogenically cooled telescope of 68.5 cm aperture diameter and two scientific 
instruments, the Far-Infrared Surveyor (FIS; \cite{FIS-Ins}) and the Infrared Camera 
(IRC; \cite{IRC-Ins}). The wide fields of view ($\sim$ 10 arcmin) covered by the large-format 
arrays in these instruments makes them highly suitable for efficient surveys.
AKARI has the capability to make pointed observations in addition 
to the All-Sky Survey, although AKARI is not a fully 
observatory-type mission in the same guise as the
Infrared Space Observatory (ISO; \cite{ISO}) and the Spitzer Space Telescope 
(\cite{SST}), due to the nature of its low-Earth Sun-synchronous orbit.

AKARI has operated normally since the launch, and is now generating large amounts 
of high-quality data on infrared sources ranging from nearby solar system objects 
to galaxies at the cosmological distances.

This paper describes the overview of the design, operation and observation of AKARI. 
Details of the scientific 
instruments and initial astronomical results based on the data mainly taken in the 
performance-verification phase 
(the first month after the opening of the aperture lid) are given in companion papers in this special 
issue.


\begin{figure}
  \begin{center}
    \FigureFile(170mm,111mm){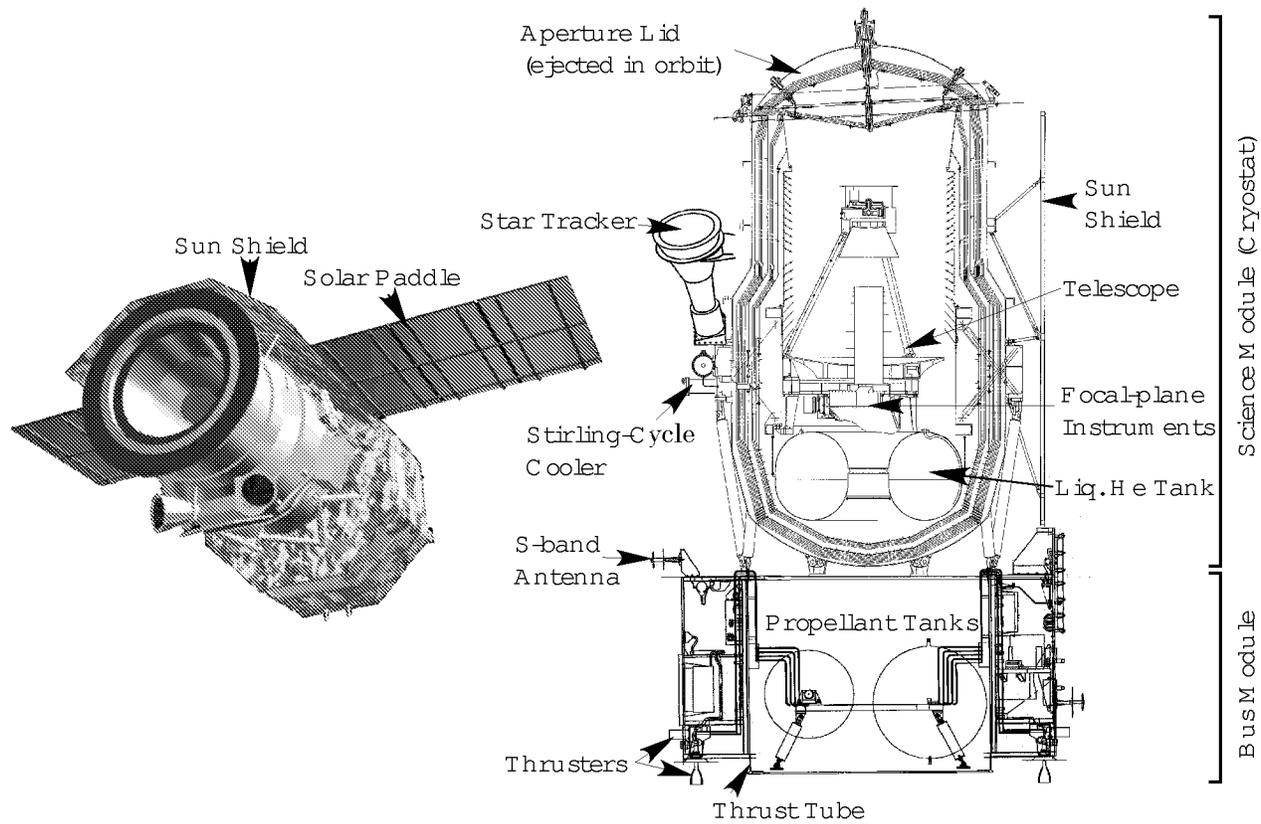}
  \end{center}
  \caption{Illustration of AKARI in orbit (left panel), and the sectional view 
(right panel).}\label{fig:AKARI_sat}
\end{figure}

\section{AKARI Satellite}

The AKARI satellite consists of two main sections, the satellite bus module and the science 
module as shown in figure \ref{fig:AKARI_sat}. The science module is a cryostat 
which contains the telescope and focal-plane instruments. The cryostat with a 
sun shield is mounted on the bus module through carbon-fiber reinforced plastic (CFRP) 
trusses, and is thereby thermally isolated from the bus. 
The satellite bus module includes the subsystems which provide functions such as the 
power supply, communications, command and data handling, attitude and orbit control, 
and temperature control and monitoring.  The key structure of the bus module is the 
cylindrical thrust tube (1 m high and 1.2 m diameter) also made of CFRP. The propellant 
tanks of the reaction control system are stored inside this thrust tube. Subsystems 
of the bus module are installed on eight instrument panels, and the panels are integrated together 
around the thrust tube. The lower end of the thrust tube is connected to top of the 
third stage of the M-V rocket. Two solar paddles are secured around the bus module in the 
launch configuration, and are deployed in orbit. Major features of AKARI are 
summarized in table \ref{tab:first}.

\begin{table}
  \caption{Design features of the AKARI satellite.}\label{tab:first}
  \begin{center}
    \begin{tabular}{ll}
      \hline
      Size & diameter 2.0 m max., hight 3.7 m (launching configuration)\\
		& width 5.5 m, hight 3.3 m (observation configuration in orbit) \\
      Mass & 952 kg in the launch configuration \\
      Orbit & Sun-synchronous polar orbit, altitude 700 km \\
      Downlink rate & 4 Mbps for scientific data \\
      Data generation rate & approximately 2 GBytes/day \\
      Data recorder capacity & 2 GBytes \\
      \hline
    \end{tabular}
  \end{center}
\end{table}

\section{Scientific instruments}

\subsection{Cryogenics}

A liquid-Helium cryostat of very high efficiency, providing a long cryogenic 
lifetime with 
a small amount of liquid Helium, was specifically designed for AKARI (\cite{Cryo}).  
This small Helium tank can provide enough room for a large aperture telescope 
in the cryostat within 
the weight and volume limits imposed by the launch vehicle. The amount of cryogen 
is only 179 liters at launch and the 
expected hold-time of the liquid Helium in orbit is
about one and a half years. 
This high efficiency has been realized by utilizing mechanical cryocoolers and 
efficient radiative cooling.  
The cryocoolers on board AKARI are two-stage Stirling-cycle coolers \citep{narasaki_04}. 
The outer shell of the cryostat is shaded from the sunlight by the sun-shield, 
and is cooled down to about 200 K 
by radiative cooling.

\subsection{Telescope}

The AKARI telescope system is a Ritchey-Chretien type with an effective 
aperture size of 68.5 cm and an f/6 system (\cite{Tel_2}, \cite{Tel}). 
Its focal plane is shared between two infrared instruments and focal-plane 
star sensors, it has a clear field of view of 38 arcmin in radius. 
The mirror material is sandwich-type Silicon Carbide (SiC), which consists of 
a porous SiC core coated with CVD (Chemical Vapor Deposition) SiC. 
The high stiffness of SiC enables us to make very light-weight mirrors. 
The primary mirror, which has a physical diameter of 71 cm, weighs only 11 kg. 

\begin{figure}
  \begin{center}
    \FigureFile(124mm,52mm){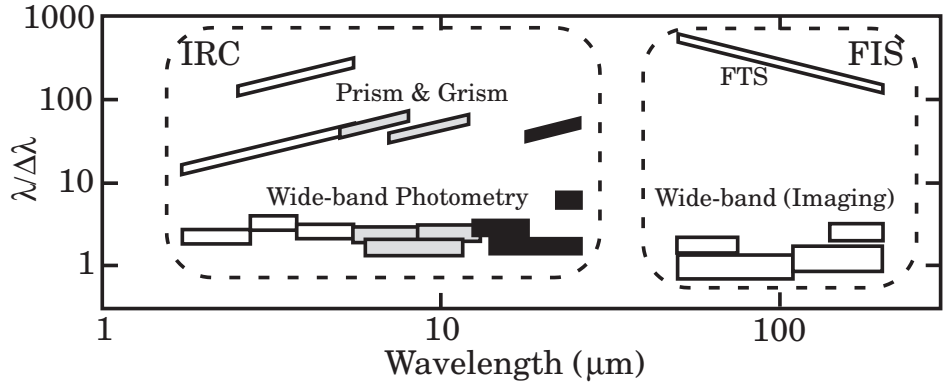}
  \end{center}
  \caption{Wavelength coverage and the resolution of IRC and FIS.}\label{fig:bands}
\end{figure}

\begin{figure}
  \begin{center}
    \FigureFile(119mm,125mm){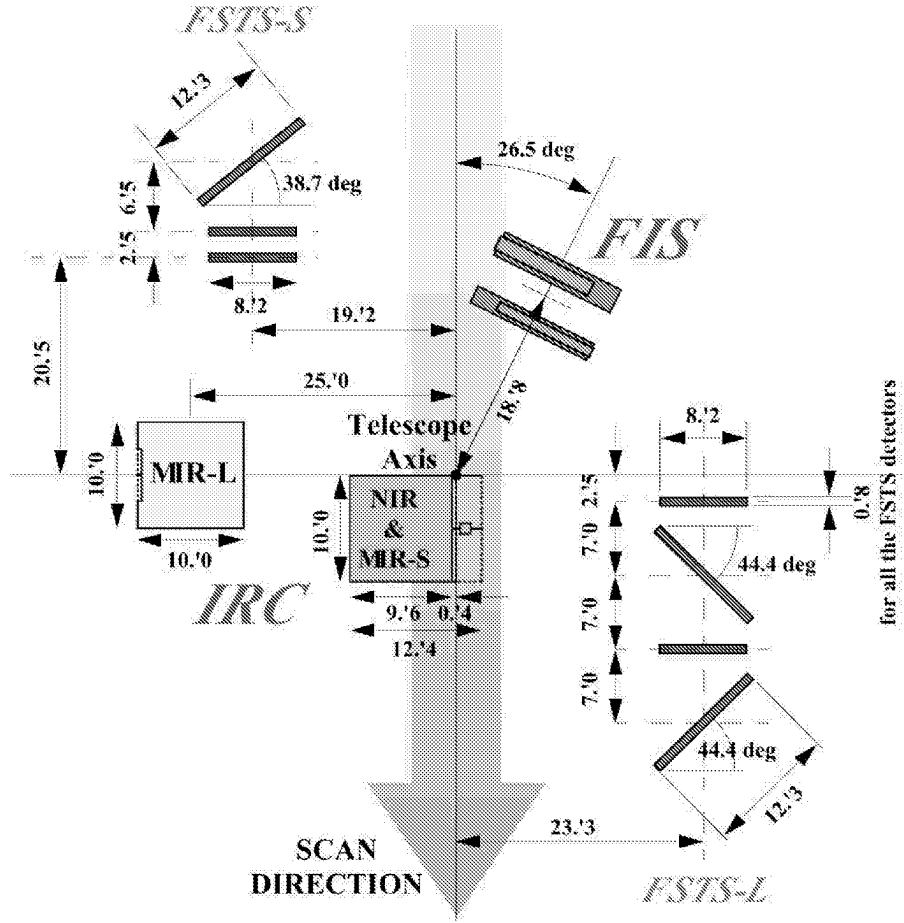}
  \end{center}
  \caption{AKARI Focal-plane layout. This figure shows a projection onto the sky. 
FSTS-S and FSTS-L are focal-plane star sensors. The scan direction in the All-Sky Survey is 
in a sense that, in this figure, FoV moves downward on the sky and stars go upward.}\label{fig:fp}
\end{figure}

\subsection{Focal-plane instruments}

One of the focal-plane instruments, FIS \citep{FIS-Ins}, is designed to perform an All-Sky Survey 
in four far-infrared wavelength bands using Ge:Ga and stressed Ge:Ga detector arrays. 
This instrument also has a spectroscopic capability via a Fourier transform spectrometer. 
The other instrument, IRC \citep{IRC-Ins}, consists of three channels, NIR, MIR-S, and MIR-L, which 
cover the 1.8--5.5\,$\mu$m, 4.6--13.4\,$\mu$m, and 12.6--26.5\,$\mu$m wavelength range, respectively. 
Each channel has three broad-band filters and additional dispersive elements. the FIS and 
IRC instruments can be operated simultaneously.

The IRC was originally designed to perform imaging and spectroscopic observations with large-format 
array detectors, pointing the telescope to a given object. However, the additional acceptable 
performance of continuous
survey-type observations with two rows of the array was confirmed in the ground tests \citep{ishihara_06}, 
and the All-Sky Survey in S9W and L18W bands 
were subsequently added to the operation modes. 

The wavelength coverage and the spectral resolution of FIS and IRC are shown in 
figure~\ref{fig:bands}, while Figure~\ref{fig:fp} shows the focal-plane layout. A brief summary of 
AKARI's scientific 
instruments is given in table \ref{tab:second}. In addition to the two scientific instruments, 
AKARI is also 
equipped with focal-plane star sensors (referred to as FSTS-S and FSTS-L), 
which are used to determine the 
telescope boresight during the All-Sky Survey.

\begin{table}
  \caption{AKARI scientific instruments.}\label{tab:second}
  \begin{center}
    \begin{tabular}{ll}
      \hline
      Cryogenics & Liquid-Helium cryostat with Stirling-cycle coolers \\
        & 179-liter super-fluid liquid Helium \\
      Telescope & Ritchey-Chretien type optics \\
        & Effective aperture 68.5 cm, total f/6 system \\
        & SiC light-weight telescope \\
      Far-Infrared Surveyor & All-Sky Survey, Imaging and Spectroscopy with FTS \\
       (FIS) & Bands: N60 (65 \micron ), Wide-S (90 \micron ), Wide-L (140 \micron ), N160 (160 $\mu$m) \\
        & Detectors: 20 $\times$ 2 \& 20 $\times$ 3 Ge:Ga arrays for N60 and Wilde-S bands \\
        & \hspace{4em} 15 $\times$ 3 \& 15 $\times$ 2 stressed Ge:Ga arrays for Wide-L, N160 bands \\
        & Pixel pitch: 29.5 arcsec for N60 and Wide-S bands, \\
		& \hspace{4em} 49.1 arcsec for Wide-L and N160 bands\\
        & Resolution for Spectroscopy: $\Delta\nu$=0.19 $cm^{-1}$ \\
      Infrared Camera & All-Sky Survey, Imaging and Spectroscopy with grisms and a prism \\
       (IRC) & Photometric Bands: NIR:  N2 (2.4 \micron ), N3 (3.2 \micron ), N4 (4.1 \micron ) \\
			& \hspace{4em} MIR-S: S7 (7.0 \micron ), S9W (9.0 \micron ), S11 (11.0 \micron ) \\
			& \hspace{4em} MIR-L: L15 (15.0 \micron ), L18W (18.0 \micron ), L24 (24.0 \micron )\\
             & Detectors: InSb 512 $\times$ 412 array for NIR, \\
			& \hspace{4em} two 256 $\times$ 256 Si:As arrays for MIR-S and MIR-L \\
             & Pixel scale: 1.46$\times$1.46 arcsec for NIR, 2.34$\times$2.34 arcsec for MIR-S, \\
			& \hspace{4em} and 2.51$\times$2.39 arcsec for MIR-S \\
			& Effective pixel scale in the All-Sky Survey: 10 arcsec (4 pixels are binned.) \\
             & Resolution for Spectroscopy: $\Delta\lambda$=0.0097 -- 0.17 $\mu$m \\
      \hline
    \end{tabular}
  \end{center}
\end{table}

\section{Satellite operations}

AKARI was initially launched into an elliptical orbit by the M-V rocket. 
The reaction control system then drove 
up the perigee altitude to bring the satellite to the observing orbit, a circular 
Sun-synchronous polar orbit at an altitude of approximately 700 km and inclination of 98.2 deg. AKARI 
flies along the day-night border 
with an orbital period of approximately 100 min. This orbit is similar 
to that of the previous IRAS satellite, and is the most suitable orbit for scanning the sky 
while keeping the telescope direction away from  the Sun and the Earth whose 
strong emission would be ruinous for the cooled telescope. 

\begin{table}
  \caption{Major events in the AKARI operation timeline.}\label{tab:third}
  \begin{center}
    \begin{tabular}{ll}
      \hline
      Events & Time (UT) \\
      \hline
      Launch & 2006 February 21 21:28:00 \\
      Injection into initial orbit & 2006 February 21 21:36:39 \\
      (Satellite separation) & \\
      Completion of orbit change maneuver & 2006 March 4 08:39 \\ 
      to the observation orbit & \\
      Aperture lid ejection & 2006 April 13 07:55 \\
      (Start of performance-verification phase) &  \\
      Start of Phase 1 observation & 2006 May 8 \\
      Start of Phase 2 observation & 2006 November 10 \\
      \hline
    \end{tabular}
  \end{center}
\end{table}

Just after the launch, it was found that the Sun aspect sensors could not detect the Sun properly.
The cause of this problem is still unknown to date. This problem
forced us to rewrite the on-board software for the attitude and orbit control subsystem, 
and delayed the opening of the aperture lid by a month. The aperture lid was finally opened on 
2006 April 13, after which AKARI began to observe the sky.  In the performance-verification phase, 
one month after the 
aperture lid opening, tuning of the scientific instruments and the attitude and orbit control subsystem, 
and telescope focus adjustment were performed. AKARI started the all-Sky Survey on 2006 May 8. 
Major events in the AKARI operation timeline are summarized in table \ref{tab:third}.

\begin{figure}
  \begin{center}
    \FigureFile(145mm,74mm){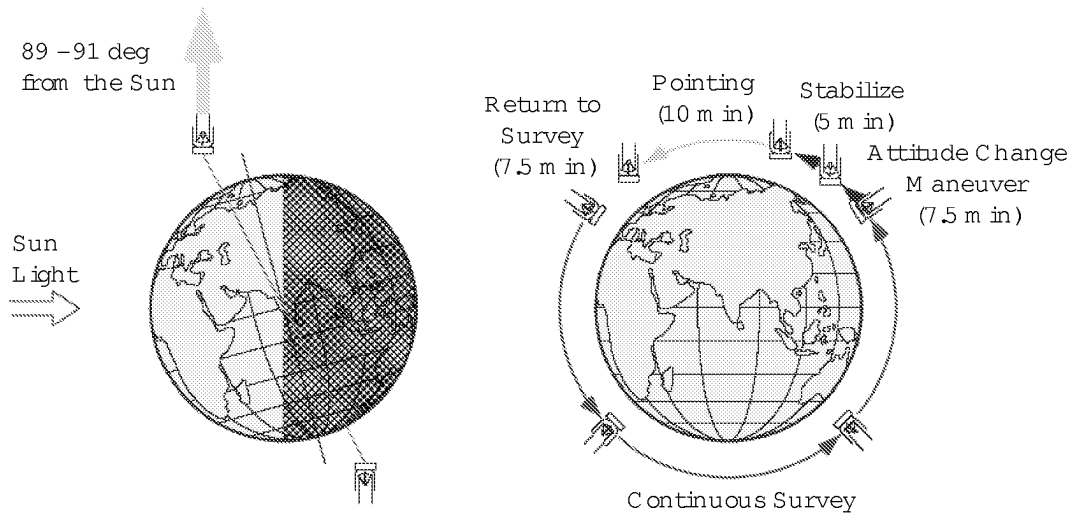}
  \end{center}
  \caption{Attitude control for observations.}\label{fig:scan}
\end{figure}

The attitude of AKARI observations are controlled as follows:  during the All-Sky Survey, 
the spacecraft rotates around the axis directed toward the Sun once every orbital revolution, avoiding 
the Earth. This results in a continuous scan of the sky at a scan speed of 3.6 arcmin/s 
(figure~\ref{fig:scan}). The whole sky can in principle be covered in half a year. The FIS and the 
IRC are also operated in a pointing mode, where the instruments observe a certain sky position for a 
longer exposure (approximately 10 minutes for one pointing with a maneuvering time of 20 minutes).

The attitude control system provides additional capabilities to shift the pointing direction by small 
amounts during the pointed observations, i.e. micro and slow scans. 
In the micro scan, the pointing direction is shifted by 
less than 30 arcsec for the purpose of dithering the IRC images. 
The slow scan 
is a continuous scan at a much slower scan speed (4-30 arcsec/s) compared to the All-Sky Survey. 
This is used to obtain sky images to significantly higher sensitivities than the All-Sky Survey.

The communications subsystem provides the command uplink in the S band, and telemetry downlink in the 
S and 
X bands. The commands are uplinked from the JAXA Uchinoura Space Center. The telemetry data are 
stored in the onboard data recorder which has a 2 GB memory, and then transmitted to the ground. 
The S-band 
telemetry normally includes low-rate engineering housekeeping data, while the high-rate (4 Mbps) X-band 
telemetry 
is used for scientific data transmission. The telemetry data are received at Uchinoura station, 
ESA's Kiruna station and also the KSAT Svalbard station. The AKARI data amounts to approximately 2 GB per 
day.

\section{In-orbit performance}

The scientific instruments are all operating normally in orbit. The temperature of the telescope and 
the IRC structure is 
5.8 K, and the temperature of the FIS detectors and the structure is 2 K or lower. 
Measurements of the Helium content performed 
in orbit has shown that 
the expected hold-time of the liquid Helium in orbit is longer than 500 days \citep{Cryo}, 
which means that the All-Sky Survey can be executed more than twice within the cryogen lifetime. 
The telescope  has a diffraction-limited performance for wavelengths longer than 7.3 $\mu$m \citep{Tel}.
The telescope pointing error is less than 3 arcsec. The attitude stability in the pointing mode 
is approximately 1 arcsec, 
and the rate stability in the All-Sky Survey is less than $10^{-4}$ deg/s. 
These numbers meet the scientific specifications for the requirements of the mission. 

The point-source flux detection limits at S/N $>$ 5 for one scan in the All-Sky Survey
are 0.05, 0.13, 2.4, 0.55, 1.4, and 6.3 Jy 
for S9W, L18W, N60, Wide-S, Wide-L, and N160 bands, 
respectively. 
These were estimated on the basis of the noise measured in orbit
using the preliminary version of the pipeline software
and could be improved with upgraded data reduction techniques. 
The chief advantage of the AKARI survey over the IRAS survey
will be the wide spectral coverage and the higher spatial resolution.
The detection limits in the two mid-infrared bands are much better than those of IRAS.  
In the far-infrared bands, the higher spatial resolution of
AKARI is expected to improve source detection and flux estimations significantly particularly in 
confused regions \citep{jeong}. 

More details on the in-orbit performance of the focal-plane instruments
are described by \citet{FIS-Ins}, \citet{IRC-Ins}, and \citet{ohyama_07}.

\section{Observation strategy}

The AKARI observations are classified into three categories, Large-Area Surveys (\cite{LS}, \cite{NEP}), 
Mission Programs (MP), and Open-Time programs (OT). The Large-Area Survey of central importance is of 
course 
the All-Sky Survey. The field of view is 8 arcmin for the FIS and 10 arcmin for the IRC.
Successive 
sky scans cover the same sky area at least twice, and enable efficient confirmation of the detection of 
celestial sources excluding false signals due to cosmic ray hits and sources of noise. The achieved
sky coverage is greater than 90 $\%$ of the whole sky during the first year,
although some areas are left unobserved or observed only once due 
to the Moon interference and disturbance by charged particles in the South Atlantic anomaly. 

In addition to the All-Sky Survey, we are also conducting
two further Large-Area Survey programs, consisting of a survey of the North Ecliptic Pole region 
(NEP) and the 
Large Magellanic Cloud. These two regions are covered with the pointed observations. 
Both are located at high ecliptic latitudes, where the density of scan paths for the All-Sky Survey 
is high and thus some observing time can be spared for pointed observations. 
Approximately  25~$\%$  of the total available pointed observations for AKARI are for use in 
the Large-Area Survey programs.

The Mission Programs are organized to interweave a series of pointed observations.
Fifteen programs on solar-system objects, star-forming regions, stars, 
interstellar matter, infrared galaxies and cosmology are being executed. 
About 45 $\%$ of the total pointed observations for AKARI are assigned to the Mission Programs.

In addition to the above observation programs, 30 $\%$ of the pointed observations 
in Phase~2 (see below) of the mission are opened to
the  Japanese, Korean, and European astronomical community.

Lastly, some pointed-observation opportunities are reserved for the calibration of instruments and 
Directors discretionary observations.

The observation periods are separated into three phases. The Phase~1 observations were made in the 
first six months after the performance-verification phase. 
AKARI performed the first All-Sky Survey during this phase and also some pointed observations at
 high ecliptic latitudes. The actual period of Phase~1 began on 2006 May 8 and ended six months later on
2006 November 9. Approximately 70 $\%$ of the sky has been covered with two or more scans 
in this period. In addition, a part of the Large-Area Surveys in 
the North Ecliptic Pole region and the Large Magellanic Cloud, were also executed.
The Phase~2 period began on 2006 November 10, 
and will last until all the Helium is exhausted. 
The second All-Sky Survey to increase the sky coverage, 
and the pointed observations for the Mission Programs are being executed during this phase. 
The Phase~3 observations are defined as those after the Helium is exhausted. 
In Phase~3, only pointed observations using the IRC/NIR channel are possible.

The point source catalogues of the All-Sky Survey are planned for release to 
the astronomical community in a timely fashion after the end of the survey.


\section{Summary}

The AKARI mission is operating normally and has been generating 2 GB of data every day since May 2006. 
Its All-Sky Survey will provide new infrared source catalogs which are expected to surpass the 
IRAS catalogs 
with higher spatial resolutions and wider spectral coverage. The AKARI mission will provide an
important and valuable database 
for the present and future research in galaxy evolution, star formation, and planet formation. 

\section*{Acknowledgements}

The AKARI project, formerly known as ASTRO-F, is managed and operated by
the Institute of Space and Astronautical Science (ISAS), 
Japan Aerospace Exploration Agency (JAXA), with the participation of 
universities and research institutes in Japan,
the European Space Agency (ESA), 
the IOSG (Imperial College, UK, Open University, UK, University of Sussex, UK, and University of 
Groningen, Netherlands)
Consortium, 
and Seoul National University, Korea.
The FIS instrument is developed by Nagoya University, ISAS/JAXA, the University of Tokyo, and the 
National Astronomical Observatory of Japan and other institutes, with contributions of NICT to the 
development of the detectors. The IRC instrument is developed by ISAS/JAXA and the University of Tokyo 
and other supporting institutes.
ESA/ESAC provides support for the All-Sky Survey data processing, 
through the pointing reconstruction.
ESAC also provides user support for the observing opportunities distributed to European astronomers. 
ESA/ESOC is providing the mission with ground support through its ground station in Kiruna.

We owe the success of AKARI 
to the dedication of many people. 
Especially, researchers of the engineering section of ISAS/JAXA have very much contributed to the
development of the AKARI satellite system. Here we list their names to express our gratitude: 
M.~Hashimoto, T.~Hashimoto, H.~Hatta, E.~Hirokawa, K.~Hirose, 
K.~Hori, T.~Ichikawa, T.~Ikenaga, 
Y.~Inatani,
K.~Inoue, N.~Ishii, T.~Kato, Y.~Kawakatsu, J.~Kawaguchi, 
Y.~Matogawa, K.~Minesugi, H.~Nakabe,
T.~Nakajima,  
I.~Nakatani, M.C.~Natori,
H.~Ogawa 
N.~Okuizumi, J.~Onoda, E.~Sato, H.~Saito, H.~Saito, S.~Sawai, M.~Shida, Y.~Sone, M.~Tajima,  
T.~Toda, K.T.~Uesugi, T.~Yamada, H.~Yamakawa. Z.~Yamamoto, and M.~Yoshikawa.

We also would like to thank the M-V rocket team led by Y. Morita for successfully launching the 
spacecraft into orbit.

Finally, we would like to thank the Science Advisory Committee of AKARI (N.~Arimoto, T.~Hasegawa, 
T.~Mukai, Y.~Nakada, S.~Okamura, M.~Tamura, and Y.~Taniguchi)
for their valuable guide to maximize the scientific outputs from AKARI.











\end{document}